# A Cubesat Centrifuge for Long Duration Milligravity Research

Erik Asphaug[1], Jekan Thangavelautham[1], Andrew Klesh[2,1], Aman Chandra[1], Ravi Nallapu[1], Laksh Raura[1], Mercedes Herreras-Martinez[1], and Stephen Schwartz[1]

[1] Arizona State University, Tempe AZ

[2] Jet Propulsion Laboratory, Pasadena CA

Send all correspondence to:

Erik Asphaug

Arizona State University

School of Earth and Space Exploration

PO Box 876004

Tempe, AZ 85287-6004

easphaug@asu.edu

tel: 480-727-2219

fax: 480-965-8102

## Abstract

We advocate a low-cost strategy for long-duration research into the 'milligravity' environment of asteroids, comets and small moons, where surface gravity is a vector field typically less than 1/1000 the gravity of Earth. Unlike the microgravity environment of space, there is a directionality that gives rise, over time, to strangely familiar geologic textures and landforms. In addition to advancing planetary science, and furthering technologies for hazardous asteroid mitigation and in-situ resource utilization, simplified access to long-duration milligravity offers significant potential for advancing human spaceflight, biomedicine and manufacturing. We show that a commodity 3U (10x10x34 cm$^3$) cubesat containing a laboratory of loose materials can be spun to 1 rpm = $2\pi/60$ s$^{-1}$ on its long axis, creating a centrifugal force equivalent to the surface gravity of a kilometer-sized asteroid. We describe the first flight demonstration, where small meteorite fragments will pile up to create a patch of real regolith under realistic asteroid conditions, paving the way for subsequent missions where landing and mobility technology can be flight-proven in the operational environment, in Low-Earth Orbit (LEO). The 3U design can be adapted for use onboard the International Space Station (ISS) to allow for variable gravity experiments under ambient temperature and pressure for a broader range of experiments.

## Introduction

We advocate flying small commodity cubesats (3U, 10x10x34 $cm^3$) as whole-spacecraft centrifuges, to recreate the off-world environments of asteroids, comets and small moons, the most common planetary bodies, in low-Earth orbit. Their regional geology appears vaguely familiar – dust plains, gravel piles and boulders, cliffs and landslides (Figure 1) – but their processes operate under gravitational stresses and dynamical timescales that are thousands of times different than on Earth, the Moon or Mars[1]. The magnitude of their surface gravity, ~0.01 $cm/s^2$ per 1 km radius, is sufficient to define an unambiguous 'down' direction, but subtle enough that landed operations are more like docking with loose material. This gives rise to dramatic topography. Materials and equipment can float freely on comets and asteroids for short timescales[2], as on the ISS, but after minutes to hours will end up on the surface.

Small body geology is fundamentally unknown, and therefore a hazardous environment. Touch-and-go sampling remains a cutting-edge technological feat[3], and controlled landing has never been achieved. Advanced operations are highly uncertain: what happens to low gravity regolith during mining or excavation? Does it go into orbit? Does it adhere to spacesuit material instead of settling? Can a spacecraft be anchored to embedded rocks, or will they pull free? Are landforms stable, or will exploration and mining activities disturb them catastrophically?

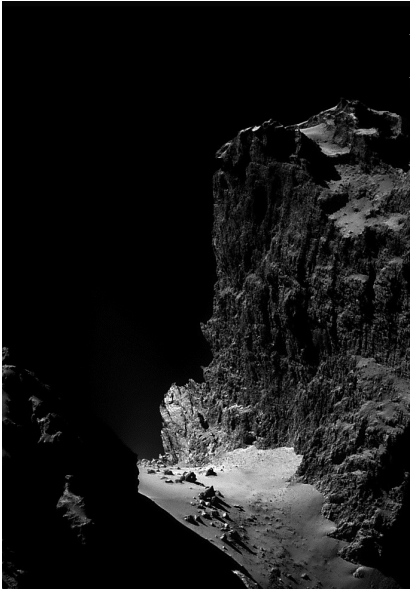 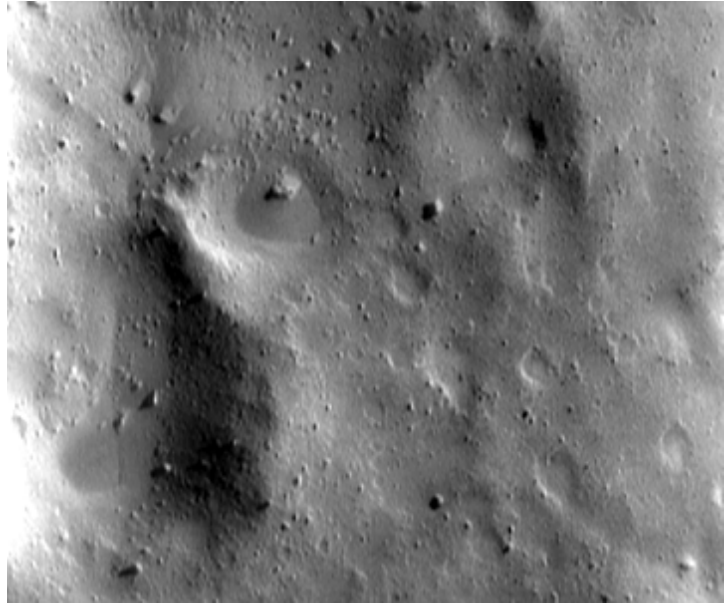

(a)

(b)

**Figure 1.** *Silicate and icy regolith in milligravity conditions.*

*(a)* *A one-kilometer cliff on the 4 km diameter comet 67P/Churyumov-Gerasimenko (C-G) imaged early in the ESA Rosetta mission (12/2014). Four-panel NAVCAM mosaic acquired from 20 km radius about the comet center. Surface gravity g~0.1 cm/s², so a leaping astronaut would land a half hour later at ~1 m/s, either into soft materials or a solid icy crust. Boulders at the cliff base are up to tens of meters diameter. Material is a combination of ices and amorphous volatiles and silicates and organics, in loose and cemented forms.*

*(b)* *Ponded and buried craters, large and small boulders, and slumps and streaks on a region of 433 Eros, a ~20 km diameter potato-shaped rocky asteroid (NASA NEAR mission, JHUAPL/Cornell). Gravity g~0.6 cm/s². Image is 600 m across, and camera pixel scale is 2 m. Material is ordinary chondrite (mostly silicate) composition, ground down by small impacts to fine sizes.*

The effect of microgravity on living organisms has been studied since the dawn of spaceflight. However we have far less knowledge whether a small but constant directional milligravity vector, imperceptible to humans on timescales of seconds to minutes, might have a cumulative effect over longer periods on biology, comparable to its pronounced effect on asteroid and comet geology. Plant germination and vegetative growth, for example[4-7], or bacterial fermentation[8] and other life processes[9-14] might operate differently under a constant directional gravity (milligravity) than under non-directional microgravity conditions.

If a small directional gravity is sufficient to overcome some of the pronounced impediments of microgravity (e.g. bone loss[13] and immune system impairment[15-17]), enabling humans and their support systems to function reliably for months or years, then a small space station with slow rotation could suffice to create milligravity conditions in Low-Earth Orbit (LEO) or in deep space, through gentle centrifugal action. Acceleration inside a centrifuge is given by $a=r\omega^2$, where $r$ is the radius and $\omega$ is the angular velocity, so a space station 10 meters across rotating once per 3 minutes would produce an acceleration equivalent to the surface gravity of asteroid Eros (Figure 1b), $g$=0.6 cm s$^{-2}$. The low rotational stresses would allow a lighter and safer spacecraft structure, compared to what is needed for Earth-like artificial gravity, and the slow rotation would minimize astronaut disorientation.

Concerning resource utilization on small bodies, milligravity conditions might represent a sweet spot in requirements and capabilities. Large ore masses could be lifted and transported at little cost of energy, while the directional gravity could be sufficient to segregate, hold, or process materials based on density, size or charge. A mining process could be optimized for asteroid-like conditions. But on small airless worlds, the challenge is not only the unfamiliar gravity. Surface particles are exposed to ionizing radiation, creating short-range forces that can vastly exceed the gravitation[18]. Pebbles and even small boulders can behave like charged polystyrene pellets (packing peanuts) on Earth – grains adhering to grains, and to surfaces. Dust might clog and damage mechanisms.

Experiments in relevant conditions are required at this juncture. Drop towers and parabolic flights can attain microgravity and milligravity conditions[19] on Earth for short durations (~1-10 s), but long-duration experiments require an accelerating frame of reference in space (a centrifuge or constant-thrusting rocket) or the surface of a small body. This leads us to consider a low cost whole-spacecraft centrifuge for creating proxy asteroid-like conditions, to enable repeated experiments in LEO, an environment that is vastly more accessible than the surface of an asteroid in deep space.

## A Whole-Spacecraft Centrifuge

The idea of a whole-spacecraft centrifuge originates with Tsiolkovsky[20] and Potočnik[21] in the early 1900s, and was popularized by von Braun[22] in the 1950s. The first demonstration was in 1966, when Gemini 11 astronauts attached a 100-foot tether between their capsule and the Agena Target Vehicle used for docking practice[23]. Thrusting against the tether, they initiated a rotation ~0.1-0.2 rpm, creating an estimated centrifugal acceleration ~0.15 cm/s$^2$, comparable to the gravity on a 10 km asteroid, that was imperceptible to either astronaut but caused a camera to slide along the instrument panel. The tethered configuration is scaleable (a 700-m tether under 1 rpm rotation would attain Mars-like gravity conditions) but in practice space tethering is a complex study in nonlinear dynamics[24].

O'Neill proposed a spinning wheel attached to a counter-rotating cylinder[25] to resolve the challenge of conserving angular momentum. Based on this approach NASA developed details for a rotating space-colony[26] in 1975. Practical efforts since then have been more modest. Japan's Centrifuge Accommodation Module (CAM) was to fly on the ISS[27] and would have enabled relatively large scale experiments from 0.01 to 2 g under ambient atmospheric conditions, where g=980 cm s$^{-1}$ is the gravity of Earth. In 2011 NASA proposed a large inflatable centrifuge[28] that would be attached to the ISS as a sleeping module, to demonstrate crewed journeys to Mars and beyond.

While large-scale and whole-spacecraft centrifuge concepts have yet to attain fruition, smaller centrifuge experiments are in operation on the ISS. KUBIK by the European Space Agency uses as a test-tube sized incubator for seeds, cells and very small animals[29], operating up to 1 g.  The European Modular Cultivation System[30] is slightly larger, 6 cm diameter, and has been used to grow plant seedlings within 1 g.  Nanorack's BioRack centrifuge[31] is of similar capability to KUBIK and can handle test-tube microbiology experiments up to 1 g.  JAXA has a small laboratory for mouse habitat experiments[32] that converts into a centrifuge operating up to 1 g, as well as the Saibo Experiment Rack consisting of the Cell Biology Experiment Facility with an incubator and small centrifuge[33].

Asteroid gravity is typically orders of magnitude smaller than these existing capabilities. Relevant experiments must contend with vibrations from spacecraft pumps and fans (typically ~0.01 cm s$^{-2}$ onboard the ISS) as well as external forces caused by spacecraft torques and tides, and air drag and turbulence. Absence of vibration is especially important for studying asteroid regolith physics, where the injection of random energy can fluidize unconsolidated materials. A free-flying centrifuge, floating inside the ISS or independently in space, is required to attain clean milligravity conditions, so we return to the idea of the whole-spacecraft centrifuge.

## A Spacecraft Proxy for Asteroids

The asteroids, comets and small moons visited to date have irregular shapes and significant expanses of regolith (Figure 1). Asteroid Eros and the Martian satellite Phobos are dust-covered bodies[34,35] about 20 km in diameter, while asteroid Itokawa, only 300 m, has centimeter-size gravels[36], its smaller grains winnowed by electrical lofting and solar wind[18]. It has been proposed[1,35] that beds of fine materials create an illusion of monolithic strength by allowing fissures to depths of 10–100 m or more, at which point gravity exceeds dry cohesion. If asteroid geology seems unknown and bizarre, the geophysics of comets is even weirder, as found out by the Rosetta mission to comet 67P Churyumov-Gerasimenko (C-G; Figure 1) during the attempted Philae landing[37].

Uncertainty as to what might happen when exploration systems interact with asteroid and comet surface materials is a serious impediment to space exploration. The misadventures of Hayabusa-1 on the surface of Itokawa[3,36], and of Philae on the surface of C-G[37], show how basic uncertainties of surface physics translate into implementation risks for flagship missions, and constrain more ambitious activities in near-Earth space and beyond. This leads us to advocate a whole-spacecraft centrifuge approach, creating patches of asteroid regolith inside of lab facilities in LEO that can be used to raise the technological readiness level (TRL) of advanced exploration systems and resource extraction technologies to TRL-9, that is, flight-proven in the operational environment.

The effective gravity of an irregular, fast-rotating asteroid or comet varies with location on the surface, even $g\sim0$ at the equator in some cases[38]. The surface potential of the 20 km diameter natural satellite Phobos, deep inside the gravity well of Mars, varies from 0.4 cm/s$^2$ at its sub-Mars point to 0.7 cm/s$^2$ at the north pole[39]. These effective gravity variations are analogous and comparable to how acceleration varies with $r$ inside a small centrifuge. So while artificial gravity is not constant inside a small centrifuge, and Coriolis effects are noticeable, this is in fact representative of actual conditions at small bodies.

## Application

How much gravity is enough, or just right, for a given artificial or natural process? How does a small but constant $g$ influence the resting configuration of rocks and airless soils? How does the presence or absence of gravity affect the operations of anchors, probes and excavators? Is a small but constant gravity of substantial benefit to humans[10-19], crop growth[4-8] and medicine[12,40-41]? In what ways is milligravity an impediment, and in what ways beneficial, to hasardous asteroid mitigation and mining? These basic questions can be answered by repeated accessible experiments in space.

A rotating cubesat can provide access to three kinds of low-gravity conditions: zero rotation (freely floating material), constant rotation (milligravity), and changing rotation (torque changing the $g$-vector, applying shear). That is the basis for the AOSAT-1 demonstration mission[42], whose science payload features optical cameras aimed at a regolith chamber, returning image data for analysis on Earth, and inertial sensors. Tunable vibrators provide additional experiments, and have the benefit of shaking granules off the viewing glass. The cubesat has a spaceflight end, roughly 1U (10cm x 10cm x 11 cm) of the chassis, and a modular lab chamber (Figure 2) with the center of mass near the 'top' of the chamber. This facilitates the separation of engineering requirements: for the spacecraft to function and return data, and for the lab chamber to run experiments and produce data. Experiments include formation of a stable pile at the angle of repose, reversal of torque to create an avalanche, and vibrators to fluidize the regolith.

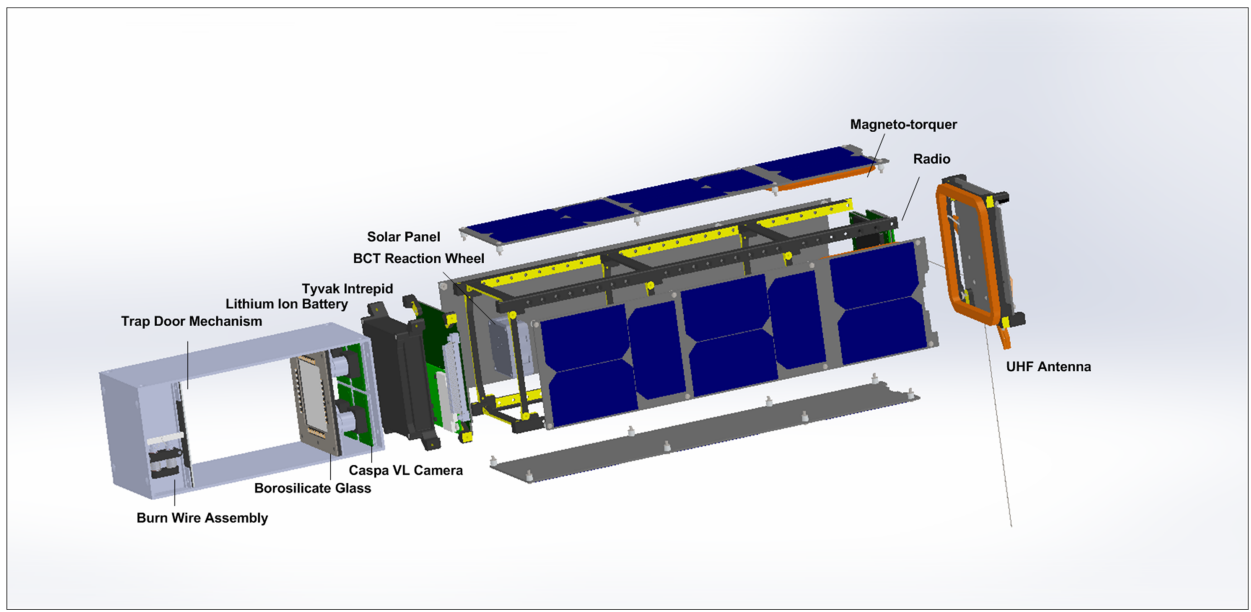

**Figure 2.** *Exploded diagram of AOSAT-1 mechanical structure[42], 10x10x34 cm (3U). The lab chamber (left) is developed and tested separately and integrated towards the end, facilitating repeated experiments. Two stereo cameras (near and far focus) are behind a glass partition, and selectable LEDs illuminate the chamber. Meteorite fragments (regolith) sieved to >3 mm are*

*released from behind a door after spacecraft deployment and systems checkout. Illustration by*

*A. Chandra.*

Experiments are conducted in a spun state, lasting minutes to hours, and communications with the ground are conducted afterwards, in a de-spun state, tracking the ground station for several orbits. Centrifuge conditions are attained using a single reaction wheel that is capable of spinning the spacecraft about its short axis (out of the plane of Figure 2) to several rpm. The wheel is sized to apply the required torque without saturation. Electromagnetic rods (magnetorquers) are used to stabilize off-axis motions during spin-up. We model this torque in combination with flywheel action and irregular spacecraft mass distribution, to show the dynamical stability of AOSAT-1 (Figure 3). Oscillations damp quickly, so that 1 rpm rotation is stabilized in 15 seconds, assuming a worst-case mass distribution (the entire regolith pile offset at a far corner of the chamber). We find that shifting the regolith mass distribution during damping has a smaller effect, so conclude that AOSAT-1 will stabilize in its experimental mode in minutes[43]. After each experiment, the magnetorquers are used to stop the rotation so that the spacecraft can point and communicate with Earth.

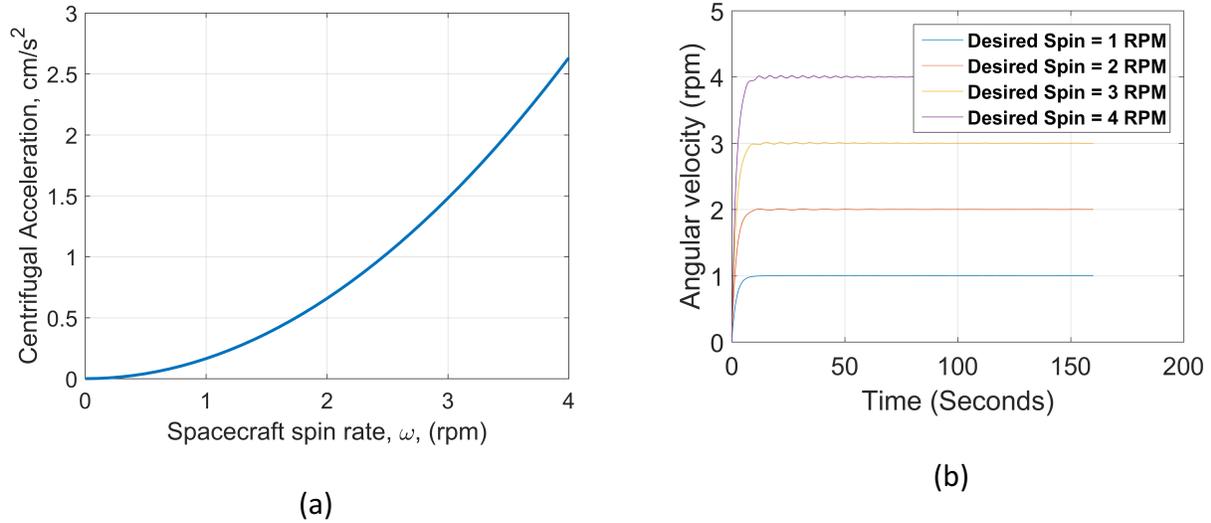

(a)

(b)

*Figure 3. Dynamical stability of AOSAT-1.*

*(a) Centrifugal acceleration (artificial milligravity) calculated as a function of angular velocity inside a 3U configuration[43], assuming r=20 cm.*

*(b) Calculated spin-up of AOSAT-1 using a single reaction wheel creates a wobble stabilized by magnetorquers, assuming a worst-case regolith distribution. Stable 1 rpm rotation ($2 \cdot 10^{-4}$ g) is obtained from a nonrotating state after 15 s, and 4 rpm ($3 \cdot 10^{-3}$ g) after ~100 s.*

AOSAT-1 experiments are conducted in vacuum. Cubesat standard allow for a pressurized laboratory up to 1.2 bar, so in principle this approach allows for similar experiments under atmospheric or nebular conditions. However, given the severe constraints on power, the laboratory temperature would have to be passively controlled. For science experiments at standard temperature and pressure it would be better to install a functionally-similar 3U chassis inside the ISS (Figure 4) with a larger motorized flywheel, spinning from 1–40 rpm to generate asteroid- to Mars-like gravity conditions. This would provode sufficient room for multiple test tubes, multiple cell cultures, or (as shown) a small plant. Unlike the free-flying cubesat, these experiments can be stopped and analyzed, replenished, restarted, and retrieved to Earth.

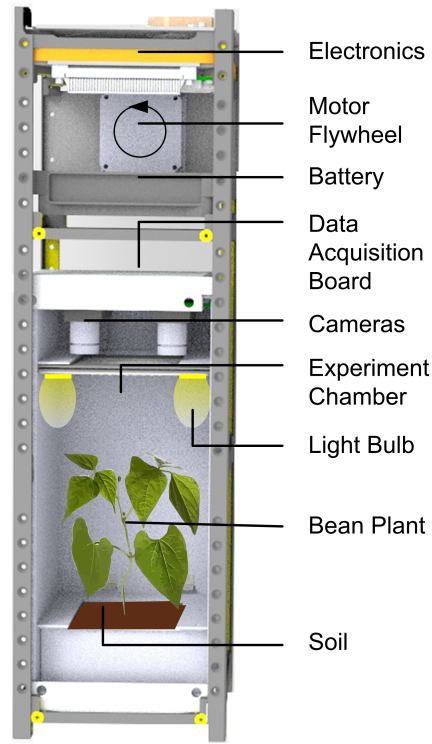

*Figure 4.* Conceptual adaptation of the 3U design, utilizing similar software and hardware, for biology and life-science experiments on board the ISS. Onboard are electronics for spin control, camera imaging, thermal and atmospheric sensing, environment control, and data acquisition. Here the lab chamber features a small plant growing under artificial light and gravity, with soil maintained at set point moisture. Illustration by A. Chandra and J. Thangavelautham.

## Discussion

The implementation of the whole-spacecraft centrifuge has waited for an affordable technology, cubesats, to meet a suitable research objective, asteroid geology. Asteroid gravity is different from microgravity[2] in that it defines a vector acceleration sufficient to create the appearance of Earth-like or lunar-like geology. For studying these worlds in accessible proxy environments, we envision using 3U cubesats to perform an increasingly detailed sequence of experiments vital to solar system research, engineering and material science. Larger centrifuges would learn from these first steps, to attain the milestone of validating full-scale asteroid landing and material transport systems without leaving LEO.

Milligravity experiments for life sciences and biology require pressurized, temperature controlled facilities; a similar 3U system could be mounted inside the ISS, utilizing common hardware, components, and software. By emphasizing commodity hardware and technology, and beginning with modest steps, these research environments can be low cost and highly accessible, thereby increasing the pace of scientific and technological advancement into the novel realm of long duration milligravity.

## Acknowledgements

We are grateful for generous advice from D.J. Scheeres (CU Boulder) and C. Hartzell (UMD) in the concept development of AOSAT, and for the efforts of three anonymous referees.

## Contributions

EA and JT conceived of the idea of a 3U centrifuge cubesat for asteroids research, and wrote the manuscript. JT led the engineering effort and AK advised the project and provided critical reviews; EA led the science definition. AC, RN, LR, MHM and SS contributed equally to system engineering, development, and project science.

## Competing Interests

The authors declare no competing interests.

## Data Availability

The data that support the findings of this study are available from the corresponding author upon reasonable request.

## Funding

Effort by EA and MHM was supported by Arizona State University, College of Liberal Arts and Sciences, Ronald Greeley Chair of Planetary Science.

# References

1. Asphaug E. Growth and evolution of asteroids. Annual Review of Earth and Planetary Sciences. 2009 May 30; 37:413-48.

2. Love SG, Pettit DR, Messenger SR (2014). Particle aggregation in microgravity: Informal experiments on the International Space Station. Meteoritics & Planetary Science 49, 732-739.

3. Yano H, Kubota T, Miyamoto H, Okada T, Scheeres D, Takagi Y, Yoshida K, Abe M, Abe S, Barnouin-Jha O, Fujiwara A (2006). Touchdown of the Hayabusa spacecraft at the Muses Sea on Itokawa. Science 312, 1350-1353.

4. Paul AL & Ferl RJ (2015) Spaceflight exploration in plant gravitational biology. (Translated from eng) Methods Mol Biol 1309:285-305 (in eng).

5. Deodelph R. The influence of simulated low-gravity environments on growth, development and metabolism of plants. Life Science Space Research. 1967;5:217-28

6. Hangarter RP. Gravity, light and plant form. Plant Cell and Environment. 1997 20:796-800

7. Kiss JZ. Mechanisms of the early phases of plant gravitropism. CRC Critical Review in Plant Sciences. 2000; 19:6: 551-573

8. Klaus DM (1998). Microgravity and its implications for fermentation biotechnology. Trends in Biotechnology 16, 369-373.

9. Souza KA, Black SD. Amphibian fertilization and development in microgravity. Physiologist. 1985 Dec 28 6:S93-4

10. Zerath E. Effects of microgravity on bone and calcium homeostasis. Advances in Space Research. 1998; 21(8-9):1049-1058

11. Loomer PM. The Impact of Microgravity on Bone Metabolism in vitro and in vivo. Critical Reviews in Oral Biology & Medicine 2001 May 1. 12(3):252 - 261

12. Blaber E, Sato K, Almedia E. Stem Cell Health and Tissue Regeneration in Microgravity. Stem Cells and Development. 2014;23:73-78

13. Vico L, Hinsenkamp M, Jones D, Marie PJ, Zallone A, Cancedda R. Osteobiology, strain, and microgravity. Part II: Studies at the tissue level. Calcified Tissue International. 2001 68(1):1-10.

14. Smith SM , Abrams SA ,Davis-Street JE, Heer M, O'Brien KO, Wastney ME, Zwart SR. Fifty Years of Human Space Travel: Implications for Bone and Calcium Research.  Annual Review of Nutrition 2014: 3(34):377-400.

15. Guéguinou N, Huin-Schohn C, Bascove M, Bueb JL, Tschirhart E, Legrand-Frossi C, Frippiat JP (2009). Could spaceflight-associated immune system weakening preclude the expansion of human presence beyond Earth's orbit? J. Leuk. Biol. 86, 1027-1038.

16. Stress Challenges and Immunity in Space, ed A. Chouker (2012) (Springer, Heidelberg).

17. Crucian B, et al. (2015) Alterations in adaptive immunity persist during long-duration spaceflight. npj Microgravity 1:15013.

18. Scheeres DJ, Hartzell CM, Sánchez P, Swift M (2010). Scaling forces to asteroid surfaces: The role of cohesion. Icarus 210, 968-984.

19. Blum J, Wurm G, Kempf S, Poppe T, Klahr H, Kozasa T, Rott M, Henning T, Dorschner J, Schräpler R, Keller HU. Growth and form of planetary seedlings: results from a microgravity aggregation experiment. Physical Review Letters. 2000 Sep 18;85(12):2426.

20. Tsiolkovsky K (1911). Investigation of Outer Space by Reaction Devices. Aeronautical Courier.

21. Potočnik, H (1929). Das Problem der Befahrung des Weltraums. Verlag Schmidt, Berlin.

22. Von Braun W (1952). Crossing the Final Frontier. Colliers, March 22, 1952.

23. Gatland, Kenneth. Manned Spacecraft, Second Revision. New York, NY, USA: MacMillan Publishing Co., Inc, pp. 180–182, ISBN 0-02-542820-9 (1976).

24. Pradhan, S., V. J. Modi, and A. K. Misra. "Tether-platform coupled control." Acta Astronautica 44, no. 5 (1999): 243-256.

25. O'Neill GK. The Colonization of Space. Physics Today, 1974, 27 (9): 32–40

26. Johnson RD, Holbrow J. Space Settlements: A Design Study. National Aeronautics and Space Administration, NASA SP-413, 1977.

27. Tsai CF, Castro H, Iwohara S, Kamiya T, Ito S, Kohama T, Kanazawa R. Centrifuge Accommodation Module (CAM) Cabin Air Temperature and Humidity Control Analysis. SAE Technical Paper; 2005 Jul 11.

28. Holderman M, Henderson E. Nautilus-X Multi-Mission Space Exploration Vehicle. NASA Johnson Space Center, Concept Presentation, pp. 1-28, 2011.

29. "KUBIK," Technical Document, European Space Agency, ESA-HSO-COU-025, 2006

30. "European Modular Cultivation Systems (EMCS)", Technical Document, European Space Agency, ESA-HSO-COU-013, 2006

31. "Space Biology Product Catalogue", Technical Document, EADS-Astrium, 2012.

32. Shimbo M, Kudo T, Hamada M, Jeon H, Imamura Y, et al., "Ground-based assessment of JAXA mouse habitat cage unit by mouse phenotypic studies," Experimental Animals 2016; 65(2): 175–187.

33. Wakatsuki T, Nishikawa W, Kobayashi R, "Preparation Status of Payload Operations for the First Experiment in JEM," Transactions of JSASS Space Technology Japan, Vol. 7, pp. th_15-20, 2009.

34. Robinson MS, Thomas PC, Veverka J, Murchie SL, Wilcox BB. The geology of 433 Eros. Meteoritics & Planetary Science. 2002 Dec 1; 37(12):1651-84.

35. Hurford TA, Asphaug E, Spitale JN, Hemingway D, Rhoden AR, Henning WG, Bills BG, Kattenhorn SA, Walker M. Tidal disruption of Phobos as the cause of surface fractures. Journal of Geophysical Research: Planets (2016).

36. Fujiwara A, Kawaguchi J, Yeomans DK, Abe M, Mukai T, Okada T, Saito J, Yano H, Yoshikawa M, Scheeres DJ, Barnouin-Jha O. The rubble-pile asteroid Itokawa as observed by Hayabusa. Science. 2006 Jun 2; 312(5778):1330-4.

37. Biele J, Ulamec S, Maibaum M, Roll R, Witte L, Jurado E, Muñoz P, Arnold W, Auster HU, Casas C, Faber C. The landing (s) of Philae and inferences about comet surface mechanical properties. Science. 2015 Jul 31; 349(6247):aaa9816.

38. Ostro SJ, Margot JL, Benner LA, Giorgini JD, Scheeres DJ, Fahnestock EG, Broschart SB, Bellerose J, Nolan MC, Magri C, Pravec P. Radar imaging of binary near-Earth asteroid (66391) 1999 KW4. Science. 2006 Nov 24; 314(5803):1276-80.

39. Shi X, Willner K, Oberst J, Ping J, Ye S (2012). Working models for the gravitational field of Phobos. Science China Physics, Mechanics and Astronomy, 55(2), 358-364.

40. Effect of Spaceflight and Spaceflight Analogue Culture on Human and Microbial Cells: Novel Insights into Disease Mechanisms (2016) (CA Nickerson, CM Ott, N Pellis, eds.). Springer Publishing, New York, NY.

41. Wilson JW, et al. (2007) Space flight alters bacterial gene expression and virulence and reveals a role for global regulator Hfq. (Translated from eng) Proc Natl Acad Sci U S A 104(41):16299-16304 (in eng).

42. Lightholder J, Thoesen A, Adamson E, Jakubowski J, Nallapu R, Smallwood S, Raura L, Klesh A, Asphaug E, Thangavelautham J (2017). Asteroid Origins Satellite (AOSAT) I: An On-orbit Centrifuge Science Laboratory. Acta Astronautica, 133, 81-94.

43. Nallapu R, Shah S, Asphaug E, Thangavelautham J (2017). Attitude control of the Asteroid Origins Satellite 1 (AOSAT 1). Advances in the Astronautical Sciences 159, 17-064.